# Excess Delay from GDP: Measurement and Causal Analysis

Ke Liu, Mark Hansen
Department of Civil and Environmental Engineering
University of California, Berkeley
Berkeley, California, United States
liuke126@berkeley.edu
mhansen@ce.berkeley.edu

*Abstract*— **Ground Delay Programs (GDPs) have been widely used to resolve excessive demand-capacity imbalances at arrival airports by shifting foreseen airborne delay to pre-departure ground delay. While offering clear safety and efficiency benefits, GDPs may also create additional delay because of imperfect execution and uncertainty in predicting arrival airport capacity. This paper presents a methodology for measuring excess delay resulting from individual GDPs and investigates factors that influence excess delay using regularized regression models. We measured excess delay for 1210 GDPs from 33 U.S. airports in 2019. On a per-restricted flight basis, the mean excess delay is 35.4 min with std of 20.6 min. In our regression analysis of the variation in excess delay, ridge regression is found to perform best. The factors affecting excess delay include time variations during gate out and taxi out for flights subject to the GDP, program rate setting and revisions, and GDP time duration.**

## I. INTRODUCTION

The Ground Delay Program (GDP), one of FAA's most commonly used Traffic Management Initiatives (TMI), has been employed in the National Airspace System (NAS) for the past three decades [1]. The purpose of GDPs is to mitigate imbalance between demand and capacity at arrival airports, by assigning delays to its incoming flights before their departure, thereby shifting expensive and unsafe airborne delay to the ground. Over time, the GDP mechanism has been refined to better serve flight operators and induce them to provide more accurate information to FAA traffic managers. The most important GDP refinement was the introduction of Ration-By-Schedule Algorithm (RBS) and Collaborative Decision Making (CDM) [2], which afforded airlines greater flexibility in assigning flight delays and improved equity. More recently, FAA has proposed the modernization of the air traffic system to improve the safety, efficiency and predictability of operations in 2025 through its NextGen program [3]. Trajectory Based Operations (TBO) and Time-Based Flow Management (TBFM) have been gradually implemented to help controllers manage traffic and reduce system uncertainties, which the agency claims will lead to a significant improvement in GDP efficiency [4]. On the other hand, despite of the significant temporarily decline in air traffic due to COVID, the system is gradually recovering [5] with traffic demand projected to grow in the range of 1.5% to 3.8% annually over the next two decades [6]. It is therefore expected that the need for TMIs, including GDPs, to balance demand and capacity in NAS will also grow.

Theoretically, an ideal GDP shifts all the necessary flight delay due to arrival capacity constraints to the ground without any excess arrival delay or throughput loss. In practice, however, GDPs may result in excess delays or fail to adequately mitigate demand-capacity imbalances on account of the substantial uncertainties in decision planning and execution phases. One recent study [4] provided the statistics on flight time uncertainties at DCA and EWR during 2019, and quantitively analyzed the impact of these uncertainties on unused airport arrival slots and excessive vectoring during GDP operations. One limitation of this study is that the factors were evaluated in isolation. Another paper analyzed the additional delay experienced by GDP-hold flights for five US airports in 2005[7]. By comparing the magnitude of delays incurred by EDCT flights and non-EDCT flights, the analysis found that gate-in delays for GDP flights are primarily caused by taxi-out delays, with minor impacts from airborne and taxi-in delay. This study emphasized the flight time variations in aircraft accuracy delivery during GDP, but method was simplistic with no strong statistic supports. Reference [8] applied a stochastic integer program and simulation model for GDP operation to evaluate the significance of demand uncertainties including flight cancellation, pop-ups, and flight arrival time deviations, and to test how the system could benefit from different arrival rate setting. The model results offered great insights on the effects of demand uncertainties and the trade-offs between airborne holding and arrival rate setting, but the model was built upon over-simplified assumptions concerning GDP procedures and distributions of demand uncertainties. Other studies focused on operational statistics to measure the performance of historical GDPs. Liu and Hansen [9] developed GDP performance metrics for airport utilization, efficiency, predictability, equity and flexibility, and applied it to SFO's GDPs in 2006 and 2011. Later research [10] adopted the same performance metrices to evaluate GDPs for GDP cluster groups based on weather forecast, GDP plan parameters and operational conditions at EWR in 2010 and 2014. It found that GDPs with low-severity weather, more revisions and large scope may have higher efficiency, while airport utilization was higher for those with high program rates, narrow scopes, and medium durations. This study provided good features in GDP assessment, but the cluster characteristics were too granular, and the results can be unstable for understanding GDP performance. A relevant work [11] studied the tradeoff of GDP equity and performance among different rationing rules by schedule, passenger, aircraft size, distance and fuel flow via an official simulator. Airline cancellation, swap, and compression were included. It concluded Ration-by-Passengers decreased significant amount of passenger delay and excess fuel burn with no flight delay increase, based on a case study on EWR in 2007. Meanwhile, a great number of research focuses on optimizing GDP design and execution. Objective functions usually involved the weighted sum of ground and air delay, and airline equity with constraints from uncertainties regarding flight delays, weather conditions and airport capacity [12-17]. Various solvers such as common machine learning, dynamic programming, stochastic optimization, and reinforcement learning, were applied to find optimal GDP parameters. Due to the great complexity in the NAS and technological limitations, such models have yet to reach the stage of practical application.

Against this background, work on GDP performance evaluation is relatively limited with relationship between its efficiency and system conditions not fully understood. To fill this gap in literature and practice, our research aims to study the factors and uncertainties in delivery accuracy of GDP operations and quantify their impacts on GDP performance. While GDP performance is multi-dimensional, we focus on one specific GDP metric, which we term "**excess delay**". Notionally, excess delay measures the additional arrival delay incurred by flights subject to a GDP, compared to a scenario in which these flights would have been allowed to depart without a GDP-assigned ground delay. To estimate excess delay, we designed a flight detection algorithm and employed a deterministic queueing model to measure excess delay from an individual GDP and applied this methodology to over 1210 GDP's that were implemented in the year 2019. The result is then used as the outcome variable for causal analysis with features generated from GDP planning and execution stages, flight time variations and airport-fixed effects.

The contributions of this research are two-fold. First, while most existing GDP performance research is limited to just one or two airports, our research applies these methodologies to a large sample of GDPs at many airports, leading to findings of greater generality and comprehensiveness. Moreover, instead of comparing flight delays of restricted and non-restricted fights, our approach involves predicting how the airport arrivals would be different in the absence of a specified GDP, and measures GDP-induced excess delay more precisely and accurately, while using widely available data sources.

Our second contribution is to determine the impacts of a wide variety of features, pertaining to rates used in planning GDPs, GDP scope and duration, and differences between planned and actual flight times on GDP excess delay. This can help identify priorities for improvement opportunities on GDPs. Specifically, our research contributes to the understanding of trade-offs between higher throughput and less airborne delay associated with changes in GDP rate calling, while also enabling assessments of the potential benefits from increasing accuracy in flight time estimation due to TBFM and TBO.

The remainder of this paper is organized as follows. Section II provides the background information on GDP. Section III presents our methodology, including GDP-restricted flight detection and delay calculation, use of a queueing model for excess delay measurement, and causal analysis on GDP excess delay. Section IV describes the data sources and relevant statistics. Model results are presented and discussed in Section V. Finally, Section VI offers conclusions discusses future directions for future research.

## II. BACKGROUND

This section introduces GDP design and execution procedure, with discussion on potential challenges in the current system.

### A.  GDP process

The GDP terminologies related this study are listed as follows,

- **Scheduled Runway Time of Arrival _(SRTA)_**: defined as Scheduled Gate Time of Arrival minus taxi time with default value of 10 minutes. Usually used to measure arrival demand and assign slots during GDP design stage.
- **ADL time**: the time when GDP event is modeled and released. GDP events including issuing, revising and cancelling GDP with notation of $ADL_0$, $ADL_i$ (i stands for $i^{th}$ revision) and $ADL_c$.
- **_Scope_**: the set of origin airports from which the departure flights to the GDP airport can be restricted.
- **GDP start/ end time**: the start/ end time of a GDP.
- **_GDP canceled time_**: the time when GDP is canceled.
- **_Program rate (PAR)_**: airport acceptance rate used in planning or revising a GDP.
- **_ARR RATE_**: airport actual acceptance rate from ASPM database.

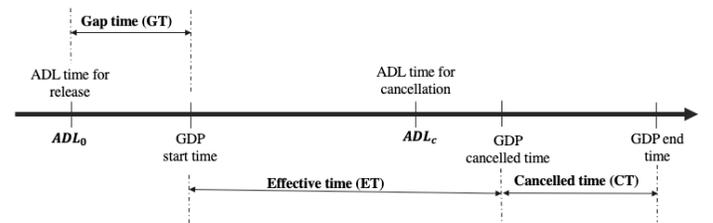

Figure 1.  GDP operation time decomposition

The motivation of GDP is to shift foreseen airborne delay due to demand-capacity imbalance at an arrival airport to the ground. In practice, Flight Schedule Monitor (FSM) is a core decision support tool used by FAA Air Traffic Control (ATC) to dynamically model, implement and monitor GDP operation [19]. A GDP is needed when arrival demand is expected to excessively exceed the airport capacity. FSM helps traffic management specialists to model and determine the best GDP parameters including start/ end time, program rate and scope, to maximize efficiency in adverse conditions. After the GDP is modeled, specialists run an RBS algorithm based on SRTA of incoming flights within the scope, to assign the arrival slots and issue each flight with a Controlled Time of Arrival (CTA) and Controlled Time of Departure (CTD), also named as Expected Clearance Time (EDCT). In FSM, CTD is calculated as,

$$CTD = CTA - ETE. \tag{1}$$

where ETE is the estimated time enroute from the flight plan. The assigned GDP delay from CTD is named as EDCT HOLD [18]. One should notice that there could be EDCT-HOLD delay not induced by GDP since other TMIs such as Air Flow Program (AFP) and Ground Stop (GS) can also result in flights with EDCT. Once the program is issued and flight control times are sent to flights, airlines can respond by cancelling or swapping some of their flights to meet their schedule needs via CDM, followed by compression to fill up open slots. GDP is somewhat flexible as program parameters can be modified to account for EDCT compliance and changes in weather or traffic conditions, and flights will be notified with updated EDCT. GDP ends when it reaches GDP end time, or the program gets cancelled [20].

## B.  Challenges in GDP

Several challenges exist in the current GDP operation, which can be divided into GDP design and execution stages. In the design phase, a GDP is a strategic TMI with a planning horizon of several hours, over which predictions of capacity and arrival demand at the GDP airport are required. However, these predictions do not adequately account for all the system uncertainties. First, the errors can come from weather forecast, so in practice, one GDP might need several revisions including program rate changes, extension, or cancellation due to changes in the weather and traffic conditions. Moreover, TFM actions are selected by ATC specialists, with guidance from FSM, in a process that relies heavily on experience and intuition [17]. Overestimation on airport called rate will assign smaller ground delay than required with the cost of larger airborne delay, while underestimation on called rate can result in conservative ground delay assignment leading to larger arrival delay and lower airport utilization. These capacity uncertainties can cause the GDP program rate and duration to differ from what is actually needed.

A second category of challenges involve accurate delivery of flights to the GDP airport. Accurate program delivery requires the restricted flights to arrive on their CTA to meet the program rate. Thus, these aircrafts are expected not only to depart within ± 5 minutes of EDCT to meet CTD, but also have airborne time close to ETE to meet CTA. However, there exists flight time variations from readiness to depart gate, taxi out time, accuracy in ETE planning by airlines and execution of ETE [4]. These deviations can result from several constraints such as aircraft maintenance, availability with aircraft and crew, surface movement constraint, terminal area constraint, weather disturbance, congested airspace, and special use of airspace [7]. Besides, pop-ups and airline flight cancellation also affect the accurate delivery in GDP. The uncertainties here can be regarded as demand uncertainties.

## III.  Methodology

In this section, we describe the methodology for estimating the excess delay from a GDP. First, key GDP times, including release, start/end, revision, and cancellation time, are determined, as shown in Figure 1. Next, we identify GDP flights and determine their GDP delays. Third, we employ a deterministic queuing model to estimate what the delay would have been in the absence of the GDP. Comparing the latter with the actual GDP delays enables us to estimate the excess delay. Finally, we develop statistical models to estimate the impact of a set of features, to be enumerated below, on excess delay, and identify those features which have the greatest impact.

## A.  Flight detection and GDP delay measurement

We define a "GDP-involved" flight as a flight which is scheduled to land at the GDP-constrained airport within initial GDP design window. We categorize such flights into three mutually exclusive groups. GDP delays are determined differently for flights in different groups.

(1) *In-scope flight (**IF**)* will satisfy following conditions:
   a)  its SRTA is within the GDP effective time window.
   b)  its departure airport is in the GDP control scope.
   c)  the flight has not taken off when the program is released.
We assume that during a GDP, in-scope flights always receive positive delay, which is to say that the assigned EDCT off time should be no earlier than flight plan off time.  If Flight Plan off is later than *GDP start time* as Figure 2 (left),

$$GDP\ delay = EDCT\ wheels\ off - Flight\ Plan\ wheels\ off$$
$$= EDCT\ Delay \tag{2}$$

If Flight Plan off time is earlier than *GDP start time* as Figure 2 (right) there is holding delay existing prior to GDP. Then,

$$GDP\ delay = EDCT\ wheels\ off - GDP\ release\ time \tag{3}$$

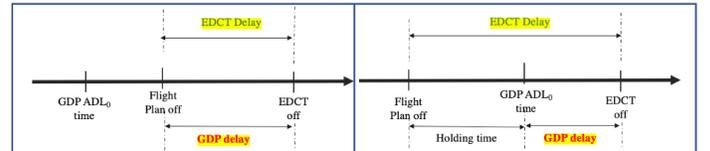

Figure 2.   GDP Delay for In-Scope flights

(2) *Cancel-delay flight (**CF**)* will satisfy following conditions:
   a)  its SRTA is within the GDP cancelled time window.
   b)  its departure airport is in the GDP control scope.
   c)  the flight plan wheels off time is earlier than ADL time for cancellation.
We assume that when a GDP gets canceled, the restricted flights on the ground can get wheels off immediately. If EDCT off time is earlier than GDP cancel time, *GDP* delay has been fully executed as Figure 3 (left). If EDCT off time is later than GDP cancel time as Figure 3 (right), GDP delay will be shorter than the initial assignment.

$$GDP\ delay = min\ (EDCT\ Delay,$$
$$GDP\ cancel\ release\ time - Flight\ Plan\ off\ ) \tag{4}$$

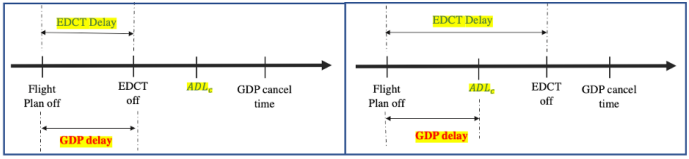

Figure 3.   GDP Delay for Cancel-delay flights

(3) *Exempt flights (EF)* will satisfy following conditions:
   a)  its SRTA is within the GDP effective time window.
   b)  its departure airport is out of restricted scope; or flight has taken off when the program is released.
Exempt flights will not receive any GDP delay.

For convenience, we use the term *GDP-restricted flights (RF)* as composite for in-scope flights and cancel-delay flights, which are flights have executed assigned GDP delays.

### B.   Queueing model

Building off our previous work on MIT performance [21], similar deterministic queuing diagram is adopted to enable the comparison between observations in actual situation and simulated condition when GDPs were removed from the system. Starting by constructing the actual scenario, we refer to actual wheels on time of flights at the given arrival airport during the study time interval $T$ (e.g., a day) to count the corresponding cumulative actual arrivals and to derive the cumulative actual arrival curve ($A(t)$). Secondly, for the simulated conditions with the absence of GDPs, model arrival curve would have behaved differently. As GDPs assign flights with only ground delays, we assume that when removing a GDP from an arrival airport, its corresponding ground delays will be eliminated as well, so the restricted flights can depart earlier by the GDP delay time. Importantly, our analysis focuses exclusively on GDP impacts, so the ground and airborne delays which are not induced by GDP should not be excluded from the computation. Hence, for a GDP restricted flight, we define the new model gate out time as actual gate out time minus GDP delay, and directly take the values from actual taxi and enroute time as model taxi out and enroute time for the same flight. Then, the model planned wheels on time for flight to enter terminal area and join the queue to wait for a runway if arrival airport is congested, can be derived as follows,

*model planned wheels on time = (actual gate out – GDP delay) +actual taxi out + actual airborne time*       (5)

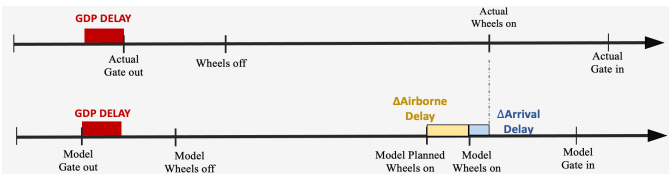

Figure 4.   Flight timeline for the actual and model scenarios

Figure 4 visualizes Eq (5) for a GDP restricted flight. The upper timeline is the actual flight time with GDP delay, while the bottom timeline is the model scenario with the GDP removed from the system. This flight has a model planned wheels on time to reach the terminal earlier, but due to the airport congestion, it might need to be held by Δairborne delay until actual landing. In the end, it still arrives earlier by Δarrival delay.

Having the model planned arrival time for each flight during GDPs, we apply the queueing model with runway capacity constraints and the first-come-first-served rule to model flight wheels-on time in simulated condition. At an arrival airport, we estimate the model throughput in each 15-min interval as the minimum of the capacity and the total demand for landing in that period, including the unsatisfied demand left from the previous period and the new arrival demand from flights whose model planned arrival times are within this period. Then, we derive the cumulative model planned arrival curve ($P'(t)$) and cumulative model actual arrival curve ($A'(t)$) in deterministic queueing diagram. By comparing $A(t)$ with $A'(t)$ in Figure 5, the green hatched area indicates the *decrease on arrival delay* if GDP was removed. The *decrease on arrival delay* is the **excess delay** from GDP, by which the flights would have been arrived earlier with no GDP. The red hatched area between $P'(t)$ and $A'(t)$ in Figure 5 represents the *increase on airborne delay* with GDP removal. In the model, flight delay change over a quarter hour is measured by taking the difference between the cumulative counts and multiplying by 15 minutes.

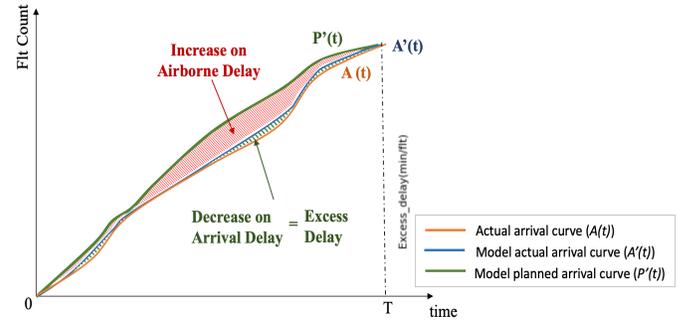

Figure 5.   Queueing diagram for actual and model scenarios

### C.   Model development for causal analysis

We develop statistical models to quantitively investigate how various factors, especially about GDP plan and execution stages, flight time variations, can affect GDP excess delay. Each observation in our data set pertains to a specific GDP. The outcome variable is the **excess delay per GDP restricted flight (min/flt)**, as measured from Section III-B. Explanatory variables are developed following the discussion in Section II-B, listed in Table 1.

One set of factors pertain to features of the GDP itself. These include effective time (**EF**), gap time (**GT**), cancelled time (**CT**) as in Figure 1, to capture the GDP planning horizon. In addition, we also include the number of revisions a GDP received during its execution (**CNT_R**). Other GDP variables include counts of the number of in-scope (**CNT_IF**), cancel-delay (**CNT_CF**), and exempt flights (**CNT_EF**). We also consider the effects of GDP scopes. As GDPs issued from the U.S. airports can assign ground delays to their arrival flights from other domestic airports and Canadian airports, our GDP scope variables separate the effects from US and Canadian airports and are defined as the average estimated time enroute ($SC\_US\_ETE$, $SC\_CA\_ETE$) for US departures and Canada departures respectively. We use the average quarter-hour rate difference between ARR RATE and annual average ARR RATE at the arrival airport during GDP ($\Delta ARR$) to measure capacity shortfall severity. Moreover, we calculate the mean and standard deviation of quarter-hour rate difference between and ARR RATE and GDP PAR ($u/\sigma PAR\_inital/\ final/\ revise$) to measure differences between the planned and actual arrival rates.

TABLE I.    DESCRIPTION OF VARIABLES

| Notation | Description |
|---|---|
| ET | Time duration of GDP execution (hr) |
| GT | Duration between GDP release and start time (hr) |
| CT | Duration between GDP cancel and plan ending time (hr) |
| CNT_R | The number of revisions during GDP execution (cancelation is not included as revision) |
| SC_US_ETE | Mean estimated time enroute (hr) from the U.S. in-scope airport to GDP airport, weighted by in-scope flight counts |
| SC_CA_ETE | Mean estimated time enroute (hr) from Canadian airport and GDP airport, weighted by in-scope flight counts |
| CNT_EF | The number of exempt flights during GDP |
| CNT_IF | The number of in-scope flights during GDP |
| CNT_CF | The number of cancel-delay flights during GDP |
| PREHOLD | Ground holding delay received by in-scope flights prior to GDP start time (hr) |
| C_SNOW/ LC/ TS/ RWY | Take value 1 if the cause of GDP is snow-ice / low ceiling/ thunderstorms/ runway construction and maintenance; take value 0 otherwise (benchmark: wind) |
| ΔARR | Rate difference between ARR RATE and airport annual average ARR RATE during GDP averaged by quarter hour |
| uPAR_initial | Mean and STD of rate difference during GDP by quarter hour: ARR RATE – initial PAR |
| σPAR_initial | |
| uPAR_final | Mean and STD of rate difference during GDP by quarter hour: ARR RATE – last-updated PAR |
| σPAR_final | |
| uPAR_revise | Mean and STD of rate difference during GDP by quarter hour: initial PAR – last-updated PAR |
| σPAR_revise | |
| ΔGO_IF | Mean and STD of gate out time difference per in-scope flight: actual gate out – EDCT gate out (min/flt) |
| σGO_IF | |
| ΔTO_IF | Mean and STD of taxi out time difference per in-scope flight: actual taxi out – unimpeded taxi out (min/flt) |
| σTO_IF | |
| ΔETE_IF | Mean and STD of enroute time difference per in-scope flight: actual enroute time – estimate time enroute (min/flt) |
| σETE_IF | |
| ΔGO_EX | Mean and STD of gate out time difference per exempt flight: actual gate out – Flight Plan gate out (min/flt) |
| σGO_EX | |
| ΔTO_EF | Mean and STD of taxi out time difference per exempt flight: actual taxi out – unimpeded taxi out (min/flt) |
| σTO_EF | |
| ΔETE_EF | Mean and STD of enroute time difference per exempt flight: actual enroute time – estimate time enroute (min/flt) |
| σETE_EF | |
| APT_BOS/ JFK/ LGA/ ORD/ PHL/ SEA/ SFO/ others | Take value 1 if GDP airport = BOS/ JFK/ LGA/ ORD/ PHL/ SEA/ SFO/ other airports; take value 0 otherwise (benchmark: EWR) |

For demand uncertainties, we consider flight time variations from gate departure, taxi out, enroute time for in-scope flights and exempt flights during a GDP. We calculate both mean and standard deviation of time difference between actual time and EDCT time (for in-scope flights) or flight plan time (for exempt flights).

To account for systematic differences in GDP arrival delay across airports, we include fixed-effect dummy variables on arrival airports to account for unobserved sources of variation at the airport level. Airports with annual GDP counts of less than 52 are grouped as "others" to address the concern that training data in cross validation should be sufficiently large and statistically representative.

For the model selection, our study aims to identify the significant factors contributing on GDP performance and to interpret results in terms of physical quantities, so three types of interpretable models — OLS, Lasso and Ridge regressions are trained and compared. After multiple data filtering steps for the designed variables, the whole dataset is randomly split into a training set (80%) and a testing set (20%). All numerical features on the training set are standardized and the testing set are rescaled with the same scaling parameters including mean and standard deviation for each feature from the training set. Then we adopt five-fold cross-validation on the training sample for hyperparameter tuning. Last, we train the models with optimal hyperparameters on the entire training set and evaluated on the testing set with root mean square error (RMSE) and mean squared error (MSE).

## IV. DATA AND STATISTICS

### A. Data and preparation

In this research, we selected the GDP operations from January 1, 2019, through December 31, 2019, for the whole NAS. Three datasets from two data sources, the TMI Advisories Database and FAA's Aviation System Performance Metrics (ASPM) are collected. These three datasets were first cleaned, filtered, and unified with time zones (UTC) before being merged into flight-based, airport-quarter-hour-based and GDP-based master datasets for performance assessment and causal analysis.

(1) *ASPM Flight Level Data*: contains records of flight plan, schedule, EDCT and actual times for individual flights arriving at 77 major airports in the U.S.

(2) *ASPM Quarter Hour Data*: operational conditions such as airport capacity (ARR_RATE), runway layout and weather at 77 major US airports, at 15-min granularity.

(3) *ATCSCC GDP Advisory Data*: include all GDP records in NAS. Each record is a GDP release, revision or cancellation event with information on scope, releasing time, start/end time, causes, program rate and cancel time.

### B. Summary statistics

Based on ATCSCC GDP Advisory Data, there were 53 airports issuing 1319 GDPs during 2019 including 4 Canadian airports (CYEG, CYUL, CYVR, CYYC, CYYZ), among which only 33 airports have records in the ASPM datasets. Therefore, our study scope covers these 33 airports which released 1210 GDPs with the annual frequency shown in Figure 6 (top). EWR (212), SFO (181) and LGA (139) issued the most GDPs. Altogether there are 8 airports with at least 52 GDPs in 2019. The bottom boxplot shows the distribution of the count on revisions a GPD received during execution by arrival airports. Around 40% of GDPs had more than one revision.

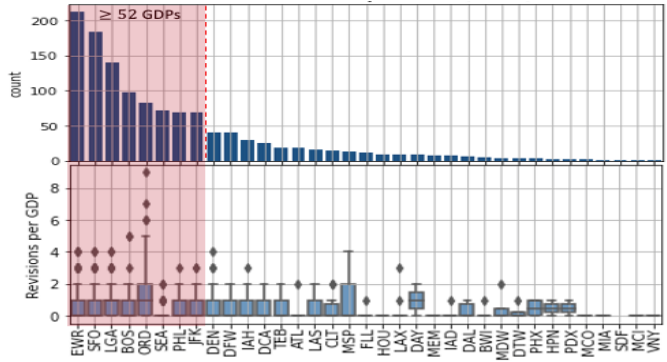

Figure 6.   GDP frequency and revisions by airports (2019)

We merge the release, revision and cancellation events for each GDP and derive their GT, ET and CT. Figure 7 (a) shows the mean time durations for top airports. On average, GDP are released 1.5hr before its start, while EWR and JFK had average GTs of close to 2hr. LGA has relatively longest effective time, and BOS canceled GDP with largest time interval. Flight detection methods are applied to all GDPs, and Figure 7 (b) presents the number of IF, EF, and CF averaged by GDP for top airports. Then, GDP delays from restricted flights are measured and grouped by airports as shown Figure 7 (c). ORD-sourced GDPs not only involved the largest average counts of IFs and EFs per program, which matches with the evidence that it led to highest average GDP delay, 346hr/GDP.

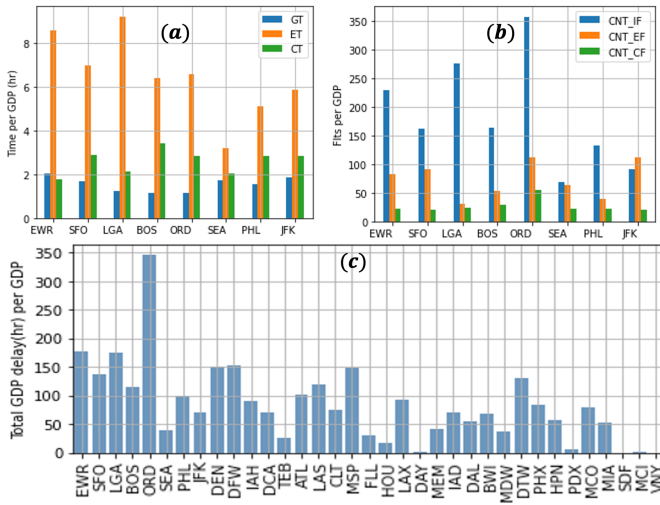

Figure 7.   GDP time, involved flights and assigned delay by aiports (2019)

## V. RESULTS AND DISCUSSION

Results are presented into two phases. In Subsection A, GDP excess delays measuring from the deterministic queueing model is analyzed. In Subsection B, after candidate model evaluation, results from the best regression models are presented.

### A.   GDP performance

We apply the queueing model to each GDPs among 33 airports during 2019 to calculate the excess delay. The bar plot in Figure 8 shows excess delay from implementing GDPs averaged by airports, ordered by their GDP issuing frequency. In addition, the average excess delay per RF is calculated by dividing excess delay by the count of restricted flights for each GDP. We use this average excess delay as the independent variable in the regression analysis. The boxplot shows the results by airports. The mean excess delay per RF over all GDPs is 35.4 min with std of 20.6 min. Among the airports with high GDP frequency, SEA is with lower excess delay while ORD and DFW with relatively higher excess with higher variance.

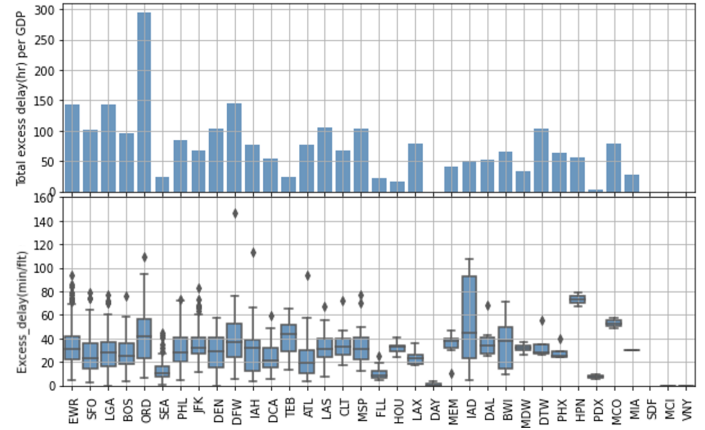

Figure 8.   Excess delay averaged by GDP(top) and RF (bottom) by airports

### B.   Statisical model results: causal analysis

Following the model specifications in Section III-C, hyperparameters for Lasso and Ridge regressions are tuned by cross validation, and the three candidate models (Lasso, Ridge and OLS) has been trained over the entire training set. We then evaluate model performances in terms of MAE and RMSE as in Table 2. MAE among the three models range from 8.9 to 9.6 min/flt and RMSE ranges from 13.4 to 14.2 min/flt. The model performances across these three models are not significantly different, but Ridge regression is slightly better. Overall, the regularization from Ridge regression enables us to avoid overly complex model by reducing model variance, and to avoid shrinking coeffects to exactly zero as Lasso does. Hence, we choose Ridge regression with hyperparameter of 0.77 as the final model for interpretation and feature analysis.

TABLE II.     MODEL PERFORMANC ON TEST DATA

|  | OLS | Lasso | Ridge |
|---|---|---|---|
| RMSE | 13.732 | 14.130 | 13.483 |
| MAE | 9.582 | 9.540 | 8.943 |

The Ridge regression results are summarized in Table 3, with estimations from Lasso and OLS added for reference. Overall, most signs of variables are consistent with our expectation based on domain knowledge.

TABLE III.   ESTIMATION RESULTS ON EXCESS DELAY PER RF (MIN/FLT)

| Var | Ridge Coef. | Lasso Coef. | OLS Coef. | P>|t| |
|---|---|---|---|---|
| const | 17.993 | 35.419 | 39.875 | 0.000*** |
| ET | 1.624 | 0.335 | 4.830 | 0.000*** |
| GT | -0.973 | -1.830 | -3.119 | 0.000*** |
| CT | 0.443 | 0.000 | 1.339 | 0.010** |
| CNT_R | 3.191 | 5.301 | 4.591 | 0.000*** |
| SC_US_ETE | 1.898 | 2.345 | 3.757 | 0.000*** |
| SC_CA_ETE | -0.312 | 0.000 | -1.188 | 0.032** |
| CNT_EF | 0.248 | 0.000 | -2.504 | 0.000*** |
| CNT_IF | 0.083 | 0.000 | -2.949 | 0.005*** |
| CNT_CF | -1.251 | -1.074 | -2.195 | 0.000*** |
| PREHOLD | 0.502 | 0.000 | -0.053 | 0.918 |
| ΔARR | -0.273 | -0.601 | -0.994 | 0.139 |
| uPAR_initial | 0.370 | 0.000 | 1.117 | 0.006*** |
| σPAR_initial | 0.363 | 0.225 | 0.072 | 0.909 |
| uPAR_final | 2.800 | 5.534 | 5.435 | 0.000*** |
| σPAR_final | 0.744 | 0.169 | 2.115 | 0.011** |
| uPAR_revise | -1.508 | -0.767 | -2.406 | 0.000*** |
| σPAR_revise | 0.586 | 0.000 | 0.415 | 0.538 |
| ΔGO_IF | -1.431 | -2.207 | -3.552 | 0.000*** |
| σGO_IF | 2.683 | 4.208 | 3.372 | 0.000*** |
| ΔTO_IF | 0.322 | 0.000 | -1.665 | 0.057* |
| σTO_IF | 1.401 | 1.244 | 3.714 | 0.000*** |
| ΔETE_IF | 0.167 | 0.000 | -0.650 | 0.326 |
| σETE_IF | 1.728 | 1.362 | 1.912 | 0.008*** |
| ΔGO_EX | 0.085 | 0.000 | -0.293 | 0.658 |
| σGO_EX | 1.706 | 2.010 | 2.717 | 0.000*** |
| ΔTO_EF | -0.155 | 0.000 | -0.111 | 0.877 |
| σTO_EF | -0.108 | 0.000 | -0.170 | 0.800 |
| ΔETE_EF | 0.127 | 0.000 | 1.624 | 0.022** |
| σETE_EF | 1.533 | 1.775 | 0.643 | 0.292 |
| C_SNOW | 4.241 | 0.000 | -7.856 | 0.000*** |
| C_LC | 2.271 | 0.000 | 0.331 | 0.916 |
| C_TS | 0.580 | 0.000 | -6.438 | 0.022** |
| C_RWY | 0.287 | 0.000 | -10.032 | 0.001*** |
| APT_BOS | 0.765 | 0.000 | -7.145 | 0.024** |
| APT_JFK | 0.982 | 0.000 | 5.013 | 0.111 |
| APT_LGA | 1.565 | 0.000 | -10.817 | 0.000*** |
| APT_ORD | 1.529 | 0.000 | 13.373 | 0.000*** |
| APT_PHL | 1.204 | 0.000 | -4.100 | 0.306 |
| APT_SEA | 0.634 | 0.000 | -7.924 | 0.044** |
| APT_SFO | 2.754 | 0.000 | -0.970 | 0.601 |
| APT_others | 4.330 | 0.000 | -1.060 | 0.694 |
| R-sqr | 0.582 | 0.514 | | 0.585 |

(* p < 0.10, ** p<0.05, *** p<0.01).

First for the GDP planning aspects, the positive coefficients of **ET** reveal that longer GDP will increase average excess delay, by 1.624 min/flt for an increase of one standard-deviation execution time. This is reasonable because a longer duration requires a longer planning horizon and greater uncertainty. **GT** coefficients are negative meaning that earlier that a GDP is released, the more flights that are eligible to be controlled for the ground delay assignment to spread the inherent risks. However, we should notice that there is a tradeoff between controlled flights and information reliability over time as there will be greater uncertainty in the weather and traffic conditions when the GDP is designed earlier. As for **CNT_R**, on average, one standard-deviation revision of GDP increases excess delay by 3.191 min/flt which is quite significant. When GDP gets revised, there are induced ground delays and active flights enroute from previous GDP setting of which the revision cannot change. Meanwhile, larger scope leads to a higher probability of realizing more unnecessary delay if the program is canceled or revised. This effect is clearly observed in **SC_US_ETE** and **SC_CA_ETE**. Both **CNT_IF** and **CNT_CF** have negative effects, as more restricted flights result in more opportunities for deviations of individual flights to offset one another.

The next variables of interest are related to PAR and ARR RATE, which accounts for the capacity uncertainties and indicates how conservative/ aggressive the GDP design is. Both **uPAR_initial** and **uPAR_final** have significant positive coefficients, implying that if the initial or last-updated PAR is smaller than ARR RATE by one standard-deviation slot per quarter-hour, the excess delay can increase by 0.37 or 2.8 min/flt respectively. Positive coefficients of **σPAR_initial** and **σPAR_final** also imply that program rate deviates more from ARR RATE will lead to more excess delay. In sum, overly pessimistic forecasts and conservative PAR can lead to excessive ground delays, as flights will be assigned with unnecessary ground delay. Additionally, **uPAR_revise** shows that if the PAR is revised to be higher than the original plan, more excess delay can be induced, as flights originating further from the airport must execute their ground delay in advance even when GDP is revised, and the accrued delays are unnecessary.

Concerning flight time variation, average time differences between EDCT/ Flight Plan time and actual flight time among in-scope flights or exempt flights during gate out (**ΔGO_IF/EF**), taxi out (**ΔTO_IF/EF**) and enroute (**ΔETE_IF/EF**) have minor impacts on excess delay. It is probably because at a macro level, under- and over-delivery of flights during the same arrival time window get balanced sometimes. However, the dispersion of time deviations from the EDCT/ Flight Plan, except for **σTO_EF**, all have significant positive impacts on excess delay. The most significant feature is **σGO_IF** as an increase of **σGO_IF** by one standard-deviation minute increases the prediction for excess delay by 2.683 min.

Additionally, we also focus on how different features contributes GDP excess delay. Feature importance of Ridge model is derived by permutation feature importance which is useful for regularized regression [22]. It compares the model score with a single feature value randomly shuffled to capture how much model depends on the selected feature, as in Eq (6).

$$\alpha_i = \frac{L\left(y, \hat{f}(X)\right) - L\left(y, \hat{f}(X_{perm\_i})\right)}{L\left(y, \hat{f}(X)\right)} \quad (6)$$

where $\hat{f}(X)$ is trained Ridge model; $y$ is excess delay per RF; $X$ is feature matrix and $X_{perm\_i}$ is new feature matrix with $i^{th}$ feature permuted; $L\left(y, \hat{f}\right)$ is error measure and we applied $R^2$ score here. Each feature is randomly shuffled for 20 times to returns an average feature importance. Figure 9 (left) visualizes the feature importance of top 25 (out of 41 in total) variables ranking by $\alpha_i$ value, while Figure 9 (right) provides the coefficients of these 25 variables from ridge regression and OLS model (with 95 confidence interval) as reference. Higher $\alpha_i$ indicates the feature contributes more to excess delay in the model. GDP final PAR setting (***uPAR_final***), GDP effective time window (***ET***), GDP revision frequency (***CNT_R***) and GDP control power (***SC_US_ETE***) have the most significant contributions on GDP excess delay, all of which are capture the feature from the supply side, while the flight time variations for in-scope flights also have relatively large effects on the model results, including ***$\sigma$GO_IF***, ***$\Delta$GO_IF*** and ***$\sigma$TO_IF***.

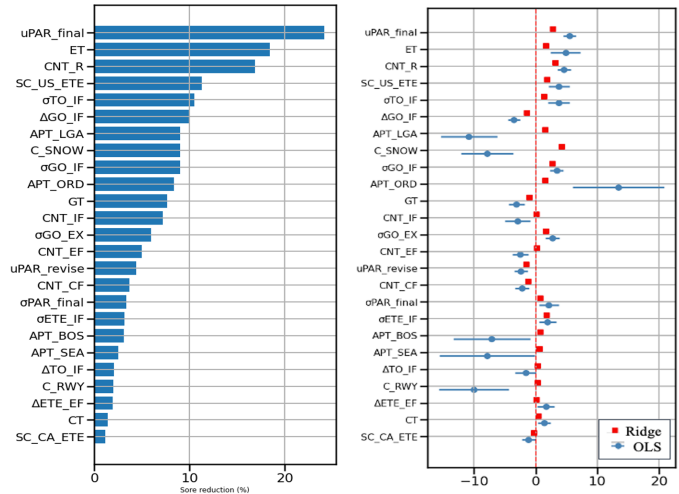

Figure 9.    Relative feautre importance for top 25 variables

## VI. CONCLUSION

In this research, we develop a methodology for measuring excess delay resulting from individual GDPs using deterministic queueing model and investigates factors that influence excess delay using regularized regression models. We applied this methodology to over 1210 GDP's that were implemented from 33 airports in the year 2019. On a per-restricted flight basis, the mean excess delay is 35.4 min per RF with std of 20.6 min. Based on model evaluation, Ridge is the best model. The model results yield several interesting insights. There is a tradeoff between control power (i.e., scope, gap time) and possibility of higher unnecessary delay. Program rate setting has significant impacts on GDP excess delay, and too conservative program rate can lead to more excessive ground delays. Uncertainties in gate out and taxi out time for in-scope flights also have great impacts on program delivery accuracy and excess delay.

The findings in this study contributes to the NAS-level evaluation on GDP excess delay, which make the analysis of greater generality and comprehensiveness. It can improve the understanding on the relationship between GDP excess delay and system conditions, especially for factors related to GDP planning and execution. It also helps identify priorities for improvement opportunities from increasing accuracy in flight time estimation and program planning. The results from regression and feature importance also reveal the importance on accurate rate setting, and the excess delay caused by revisions.

Based upon our work, future study can be extended in a few directions. First, the existing GDP delay assignment, pop-ups, airline compression, and swap behaviors can be further studied to address the demand uncertainties. Second, more statistical analysis is needed to gain insights into the dynamic decision-making procedure in GDP operation, especially about the relationship between GDP program rates and actual arrival rate. Third, the model can be complemented with one that addresses the other half of the GDP loss function—extra holding and flight time from overly optimistic assumptions about future operating conditions when planning GDPs.